\documentclass
[aps,pra,reprint,twocolumn,superscriptaddress,showpacs,showkeys,superscriptaddress,longbibliography]{revtex4-1}
\usepackage{dcolumn}    
\usepackage{ifpdf}
\usepackage{amssymb,lineno,amsfonts}
\usepackage{graphicx}   
\usepackage{dcolumn}    
\usepackage{bm}         
\usepackage{bbm}
\usepackage{mathrsfs}
\usepackage{upgreek}
\usepackage{mathtools}
\usepackage{epstopdf}
\usepackage{setspace}
\usepackage{hyperref}
\usepackage{float}
\usepackage{amsmath}
\usepackage[normalem]{ulem}
\usepackage[usenames,dvipsnames]{xcolor}
\usepackage[matrix,frame,arrow]{xypic}
\usepackage[mathscr]{euscript}

\definecolor{med-blue}{RGB}{25,25,112}
\hypersetup{colorlinks, linkcolor={red},citecolor={blue}, urlcolor={MidnightBlue}}

\newcommand{\ket}[1]{\vert{#1}\rangle}

\newcommand{\tr}{\mathrm{Tr}}

\begin{document}
	\title{Observation of interaction induced blockade and local spin freezing in a \\ NMR quantum simulator}
	\author{V. R. Krithika}
	\email{krithika$_$vr@students.iiserpune.ac.in}
\affiliation{Department of Physics, 
		Indian Institute of Science Education and Research, Pune 411008, India}
\affiliation{NMR Research Center, 
		Indian Institute of Science Education and Research, Pune 411008, India}
	\author{Soham Pal}
	\email{soham.pal@students.iiserpune.ac.in}	
	\affiliation{Department of Physics, 
		Indian Institute of Science Education and Research, Pune 411008, India}
\affiliation{NMR Research Center, 
		Indian Institute of Science Education and Research, Pune 411008, India}
\author{Rejish Nath}
\email{rejish@iiserpune.ac.in} 
\affiliation{Department of Physics, 
		Indian Institute of Science Education and Research, Pune 411008, India}
\author{ T. S. Mahesh}
	\email{mahesh.ts@iiserpune.ac.in}
	\affiliation{Department of Physics, 
		Indian Institute of Science Education and Research, Pune 411008, India}
\affiliation{NMR Research Center, 
		Indian Institute of Science Education and Research, Pune 411008, India}
	
	\begin{abstract}
{
We experimentally emulate interaction induced blockade and local spin freezing in two and three qubit Nuclear Magnetic Resonance (NMR) architecture. These phenomena are identical to the Rydberg blockade and Rydberg biased freezing.
In Rydberg blockade, the simultaneous excitation of two or more atoms is blocked due to the level shift induced by the strong Van der Waal's interaction. In such a strong interaction regime, one can also observe Rydberg biased freezing, wherein the dynamics is confined to a subspace, with the help of multiple drives with unequal amplitudes.  
Here we drive NMR qubits with specific transition-selective radio waves, while intermittently characterizing the quantum states via quantum state tomography.  This not only allows us to track the population dynamics, but also helps to probe quantum correlations, by means of quantum discord, evolving under blockade and freezing phenomena.
While, our work constitutes the first experimental simulations of these phenomena in the NMR platform, it is also the first experimental demonstration of Rydberg biased freezing.
Moreover, these studies open up interesting quantum control perspectives in exploiting the above phenomena for entanglement generation as well as subspace manipulations.}
	\end{abstract}
		
\keywords{blockade, freezing, NMR}

\maketitle

 \section{Introduction}
  \label{Introduction}
  The blockade phenomenon in which one particle prevents the flow or the excitation of other particles due to inter-particle interactions has been a subject of intense study using various quantum systems. For instance, blockade has been observed in electrons \cite{ono02}, photons \cite{bir05, lan11, hof11, van18, sni18}, ions \cite{fen16}, and Rydberg atoms \cite{gae09,urb09,bar14}. The effect of blockade has been used for the controlled preparation of quantum states \cite{pet05,now07}, in particular the entangled or non-classical states \cite{far08,pey12,mul15},  thus becoming highly relevant for quantum information applications \cite{luk01,saf10} and quantum many body physics \cite{bro20}.  In the Rydberg blockade regime, a new feature has been predicted recently by Vineesha et.al \cite{sri19}, called the Rydberg biased freezing  in which the dynamics of atoms driven with small Rabi coupling freeze. The phenomenon of biased freezing can provide local control on selected qubits, which is of vital importance in many quantum computing and information processing tasks \cite{burgarth2010scalable}.
  
In this work, we experimentally demonstrate interaction induced blockade and spin freezing, identical to the Rydberg blockade and Rydberg biased freezing, respectively, but using nuclear spins in Nuclear Magnetic Resonance (NMR) architecture. Due to the long coherence times and the ease of controlling and manipulating qubits, NMR provides an ideal platform to probe such quantum phenomena \cite{jones2000nmr,laflamme2002introduction,vandersypen2005nmr,suter2008spins,oliveira2011nmr}. Here, we demonstrate blockade and freezing phenomena using two and three-qubit NMR registers. We periodically monitor the full state of the quantum system using quantum state tomography (QST), which also allows us to estimate quantum discord, that quantifies quantum correlations in a general quantum state, pure or mixed. \cite{ollivier2001quantum,luo2008quantum,mahesh2017quantum}. 

The paper is organized as follows. In Sec. \ref{Theory}, we briefly outline the phenomena of both Rydberg blockade and Rydberg biased freezing. In Sec. \ref{Methodology}, we explain the NMR architecture, the experimental setup, and introduce the theoretical background for studying evolution of quantum correlations in the system. We then explain how blockade and freezing phenomena can be realized using NMR spin systems in Sec. \ref{Methodology} and the corresponding experimental results are discussed in Sec. \ref{NMR-Rydberg}. Finally, we conclude in Sec. \ref{Conclusions}.

\section{Rydberg Blockade and Rydberg biased Freezing}
\label{Theory}
In this section, we briefly review the phenomena of the Rydberg blockade and Rydberg biased freezing for two atoms ($N=2$). Each atom comprises of two-levels with the ground state  $\{\ket{g}\}$ coupled to the Rydberg state $\{\ket{e}\}$ by a laser field of Rabi frequency $\Omega_i$ and detuning $\Delta_i$.  In the frozen-gas limit, the system is described by the Hamiltonian ($\hbar=1$)
\begin{equation}
\hat H = -\sum_{i=1}^{2}\Delta_i\hat \sigma_{ee}^i+\sum_{i=1}^{2} \Omega_i \hat \sigma_{x}^i +V_0 \hat \sigma_{ee}^1 \hat \sigma_{ee}^2,
\label{H1}
\end{equation}
where $\hat\sigma_{ab} = |a\rangle \langle b|$ with $a,b \in\{g,e\}$, $\hat\sigma_x^i = \hat\sigma^i_{eg}+\hat\sigma^i_{ge}$. The interaction potential between two Rydberg excitations separated by a distance $R$ is given by $V_0=C_6/R^6$ where $C_6$ is the van der Waals coefficient \cite{saf10}. Henceforth, we take $\Delta_i=0$, and work in the basis $\{|gg\rangle, |ge\rangle, |eg\rangle, |ee\rangle\}$.
\begin{figure}
	\includegraphics[trim=10cm 1.5cm 7cm 1cm,width=8.5cm,clip=]{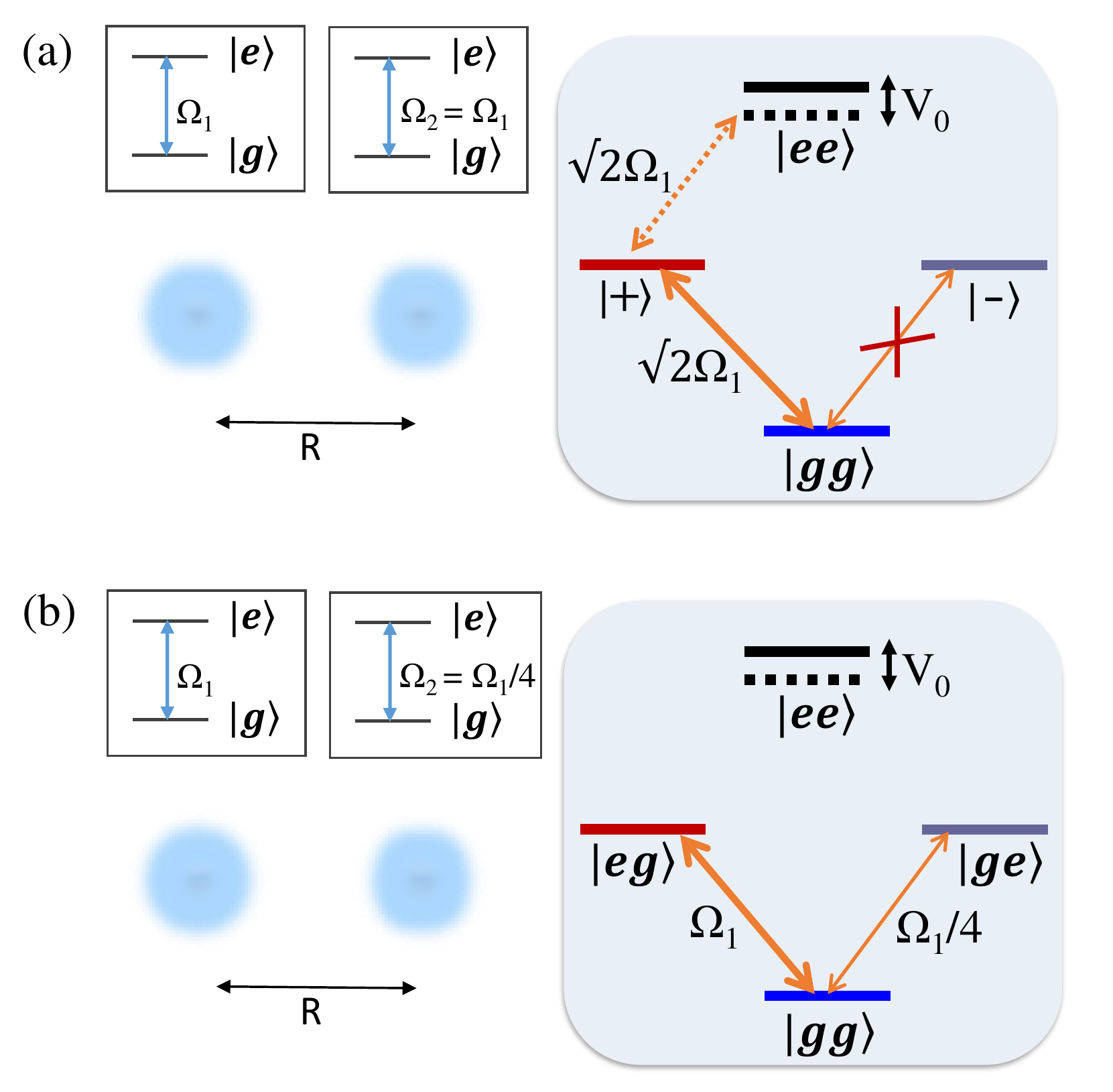}
	\caption{The energy level diagram and allowed transitions under (a) blockade and (b) Rydberg biased freezing of two interacting Rydberg atoms. The ground state of each atom is coupled to the excited state by laser fields with Rabi frequencies $\Omega_1$ and $\Omega_2$ respectively. Under strong interaction, the doubly excited state shifts out of resonance. In the regime $\Omega_1$ = $\Omega_2$ (a), the system exhibits dynamics between the ground state $\ket{gg}$ and the entangled state $\ket{+} = (\ket{eg} + \ket{ge})/\sqrt2$ with enhanced frequency $\sqrt{2}\Omega_1$. The entangled state $\ket{-} = (\ket{eg} - \ket{ge})/\sqrt2$ is also cut-off from the system's dynamics. In the regime where second atom is driven by a much weaker drive than the first atom (b), the dynamics of the second atom is suppressed and that of the first atom is unhindered, resulting in freezing of the second atom.}
	\label{schematic}
\end{figure}

\textit{Rydberg Blockade:} First we assume $\Omega_1=\Omega_2$ and for $V_0\gg\{\Omega_1,\Omega_2\}$, the doubly excited state experiences a large energy shift [see Fig.~\ref{schematic}]. In this case, if the two atoms are initialised in $|gg\rangle$, they exhibit coherent Rabi oscillations between $|gg\rangle$ and $|+\rangle = (|ge\rangle + |eg\rangle)/\sqrt{2}$ with an enhanced Rabi-frequency of $\sqrt{2}\Omega_1$, cutting off $|ee\rangle$ entirely from the population dynamics. Effectively, strong interactions hinder the presence of two excitations simultaneously, over a separation of $R_b$, the blockade radius. This phenomenon is called Rydberg blockade \cite{gae09,urb09,bar14}. 

\textit{Rydberg biased freezing:} Keeping $V_0\gg\{\Omega_1,\Omega_2\}$ (blockade regime), and increasing $\Omega_2$ eventually freezes the dynamics of the first atom. This phenomenon was first shown by Vineesha et.al \cite{sri19} and is termed as Rydberg biased freezing. Note that, the Rydberg biased freezing emerges as a combined effect of both strong interactions and the strong driving on one atom \cite{sri19}. For the strong bias on the second atom, the system exhibits coherent Rabi oscillations between $|gg\rangle$ and $|ge\rangle$, freezing the first atom. It is straightforward to extend both blockade and freezing phenomena for more than two atoms.

For $N$ atoms and $\Delta_i=0$, the Hamiltonian in Eq.~(\ref{H1}) can be extended as, 
\begin{equation}
\hat H =  \sum_{i=1}^{N} \Omega_i \hat \sigma_{x}^i +\sum_{i<j}^NV_{ij} \hat \sigma_{ee}^i \hat \sigma_{ee}^j,
\label{HN1}
\end{equation}
where $V_{ij}=C_6/r_{ij}^6$ and $r_{ij}$ is the separation between $i$th and $j$th atoms. A fully blockaded sample of $N$ two level atoms exhibit coherent Rabi oscillations between the many-body ground state $|G\rangle=\otimes_{i=1}^N|g^{(i)}\rangle$ and a collective single excited state, $|+_N\rangle=\sum_i|g g ... e^{(i)} ... g g\rangle/\sqrt{N}$ \cite{hei07}. Freezing one or more atoms in an N-atom system can also be realized by appropriately tuning the Rabi frequencies on selected atoms. In this case, the system exhibits coherent Rabi oscillations between $\ket{G}$ and the corresponding product states.

To identify both the blockade and freezing regions using an NMR setup, it is more desirable to work with the corresponding spin-model for Eq.~(\ref{HN1}). For that, we introduce the spin-1/2 operators, $\hat{I}_\alpha^i$ ($\alpha \in \{x,y,z\}$) by mapping $|g\rangle$ and $|e\rangle$ with up  ($|\uparrow\rangle$) and  down ($|\downarrow\rangle$) spin states along the $z$-axis, respectively. Then, we have $\hat\sigma_x^i=2 \hat{I}^i_x$, and $\hat\sigma_{ee}^i=(\mathbbm{1}-2\hat{I}^i_z)/2$, where $\mathbbm{1}$ is the identity operator, and the Hamiltonian in Eq.~(\ref{HN1}) reads as (apart from an identity term),
\begin{equation}
\hat H = 2\sum_{i=1}^{N} \Omega_i\hat{I}_x^i -\sum_{i=1}^{N} \bar{V}_{i} \hat{I}_z^i+\sum_{i<j}^{N} V_{ij}\hat{I}_z^i \hat{I}_z^{j},
\label{ryd_spin}
\end{equation}
where, $\bar{V}_{i} = \sum_jV_{ij}/2$. In spin models, the first term in Eq.~(\ref{ryd_spin}) plays the role of a transverse field, second term acts as a longitudinal field and the third term provides the Ising interactions. Below, we describe how to realize the above Hamiltonian using nuclear spins. For our convenience, we continue to use the states $|g\rangle$ and $|e\rangle$ as the two spin states of the NMR qubit.


\section{NMR Methodology}
\label{Methodology}
\subsection{Spin system and the Hamiltonian}
\begin{figure}[h]
	\centering
	\includegraphics[trim=6.5cm 6cm 6.3cm 1cm,clip=,width=8.5cm]{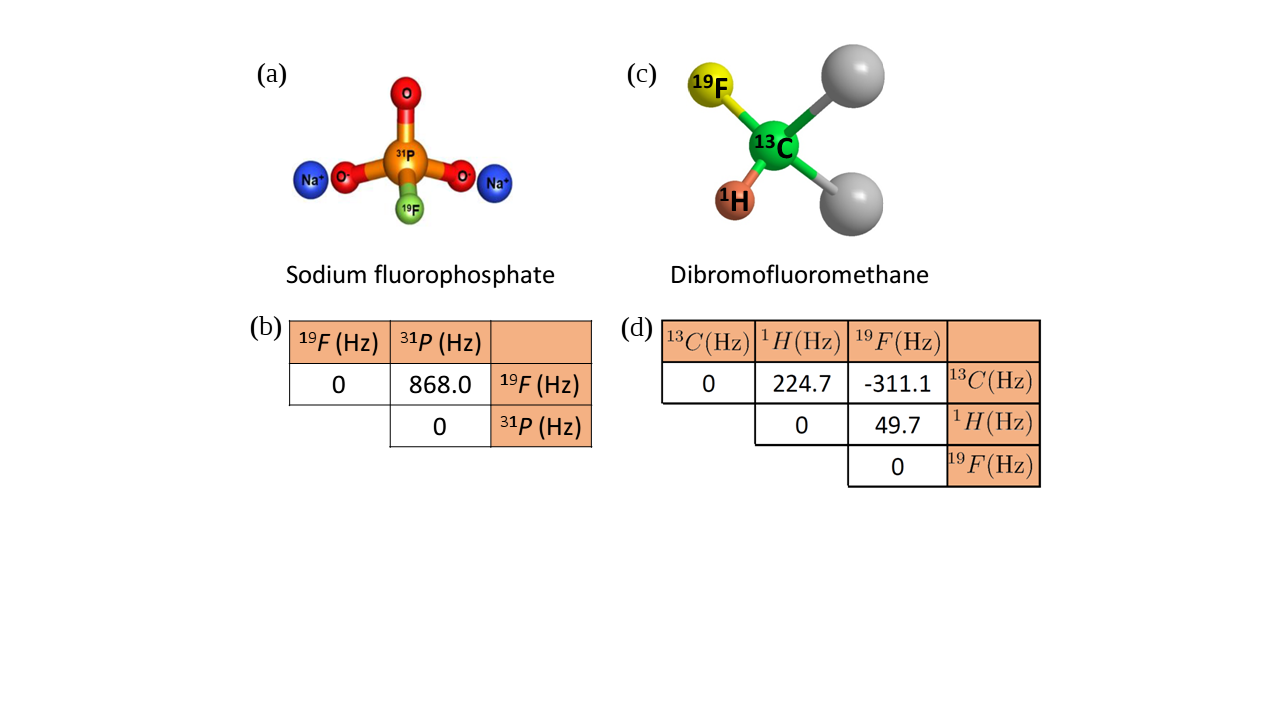}
	\caption{Sodium fluorophosphate molecule (a) and its Hamiltonian parameters (b) forming the two-qubit register.    Dibromofluoromethane molecule (c) and its Hamiltonian parameters (d) forming the three-qubit register. In tables (b,d) the diagonal elements represent tunable off-set frequencies $\nu_{i}$, and off-diagonal elements show scalar coupling constants $J_{ij}$ between the respective qubits.}
	\label{molecules}
\end{figure}

The emulations of Rydberg atom dynamics are performed on two different systems: (i) a two-qubit system involving $^{19}$F and $^{31}$P nuclear spins of sodium fluorophosphate dissolved in D\textsubscript{2}O (Fig.\ref{molecules}(a)), and (ii) a three-qubit system involving $^{1}$H, $^{13}$C and $^{19}$F nuclear spins of dibromofluoromethane, (Fig.~\ref{molecules}(c)) dissolved in deuterated acetone. All experiments were performed on a 500 MHz high-resolution Bruker NMR spectrometer at ambient temperatures. Each NMR sample contains about $10^{15}$ molecules (nuclear spin-systems) placed in an external magnetic field  $\mathbf{B}=B_0\hat{z}$, where $B_0 = 11.75$ T. The Zeeman interaction lifts the degeneracy between the spin states $m = \pm 1/2$ with an energy gap $\hbar \gamma_i B_0$  where $\gamma_i$ is the gyromagnetic ratio of the nuclear isotope and $\gamma_i B_0$ constitutes its Larmor frequency.
The time-averaged local field at the site of nuclear spins in a rapidly reorienting liquid molecule differs from the external magnetic field. The resulting individual Larmor frequencies $\gamma_i B_0(1+\delta_i)$ are strongly dependent on the chemical environment. Each of the nuclear isotopes forming our spin systems can be irradiated with circularly polarized radio-frequency (RF) waves $B^\mathrm{RF}_i \exp(i2\pi \eta_i t)$ characterized by controllable amplitudes $2\pi\nu_{i}^\mathrm{RF} = \gamma_i B^\mathrm{RF}_i$ and controllable carrier frequencies $\eta_i$.
The resonance offsets w.r.t. the carrier frequencies are given by $2\pi \nu_i = \gamma_i B_0(1+\delta_i)-2\pi \eta_i$.
The spins also interact with one another via a constant scalar coupling  $J_{ij}$ mediated through  covalent bonds. While $J_{ij}$ itself is not a controllable parameter, the effective evolution time of the scalar interaction, can however be manipulated, if required. For both the spin systems described above, the resonance offsets and coupling strengths are tabulated in Fig.~\ref{molecules}(b,d).

Thus the NMR Hamiltonian in a frame co-rotating with individual RF carriers under secular approximation is

\begin{equation}
H_{\mathrm{NMR}} =2\pi\sum_{i=1}^N \nu_{i}^\mathrm{RF} \hat{I}_x^i -2\pi\sum_{i=1}^{N}  \nu_i \hat{I}_z^i + 2\pi\sum_{i,j>i}^{N} J_{ij} \hat{I}_z^i \hat{I}_z^j.
\label{intH1}
\end{equation}
Comparing with Eq.~(\ref{ryd_spin}), we can map $2\pi\nu_{i}^\mathrm{RF}$, $2\pi\nu_i$ and $2\pi J_{ij}$ directly to $2\Omega_i$, $\bar V_i$ and $V_{ij}$, respectively. Thus, NMR systems along with RF pulses provide a natural test bed for emulating similar physics that can be studied using a Rydberg quantum simulator.

\subsection{Quantum correlation: Discord}
In addition to observing blockade and freezing dynamics in nuclear spins, we also study how quantum correlations between the qubits evolve as the system is driven under the same phenomena. We quantify these correlations using the measure of quantum discord, which is defined in terms of the mutual information in a bipartite system \cite{ollivier2001quantum}. For a given density matrix $\rho$, the information content is quantified by the von Neumann entropy $H(\rho) = -\tr(\rho \log \rho)$. 
For a bipartite system $AB$, the mutual information between $A$ and $B$ is defined as
\begin{equation}
{\cal I}(A:B) = H(A)+H(B)-H(AB)
\label{iab}
\end{equation}
where $H(A)$, $H(B)$ and $H(AB)$ are von Neumann entropies of subsystems $A$, $B$, and the entire system $AB$ respectively. The mutual information can alternatively be defined as
\begin{equation}
\mathsf{J}(A:B) = H(B)-H(B|A)
\label{jab}
\end{equation}
where $H(B|A)= \sum_{i}p_i^A H(B|A=i)$ is the entropy of subsystem $B$ conditional to a measurement on subsystem $A$ giving a result $i$ from the possible outcomes of A, with probability $p_i^A$ \cite{ollivier2001quantum}.  

The above two definitions of mutual information are classically equivalent. However, quantum mechanically this equivalence does not hold since the second definition involves measurement, which is basis dependent and changes the state of the system following a measurement \cite{katiyar2012evolution}.  The minimum difference between these two ways of evaluating mutual information quantifies a quantum correlation and is called quantum discord.
Since $\cal I(A:B)$ is independent of measurement basis,
the discord can be estimated by maximizing $\mathsf{J}(A:B)$ over all possible orthonormal measurement bases $\{\Pi_i^a\}$  on subsystem $A$.  Thus we define the  residual correlation 

\begin{equation}
D(B|A) = {\cal I}(A:B)-\underset{\{\Pi_i^a\}}{\mathrm{max}}~\mathsf{J}(A:B),
\end{equation}
as quantum discord between $A$ and $B$ \cite{ollivier2001quantum}. Note that discord is not necessarily symmetric under system partitions  and it varies from 0 for states without quantum correlations to 1 for maximally entangled states. 

\subsection{Initialization, Readout,  and Modeling Experimental Imperfections}
\label{imp}
\textit{Intialization:} At ambient temperatures, the thermal energy is much larger than the Zeeman energy gaps and accordingly an $n$-qubit NMR system is found in a highly mixed state of the form $\rho_\mathrm{th} = \mathbbm{1}/2^n + \sum_i  \epsilon_i \hat{I}_z^i$, where $\mathbbm{1}$ accounts for the uniform background population and $\epsilon_i = \hbar \gamma_i B_0/(2^n k_BT) \sim 10^{-5}$ is the purity factor capturing the deviation population distribution. Therefore, one prepares a \textit{pseudopure state} (PPS) \cite{cory1997ensemble} of the form $\rho_{\mathrm{pps}} = (1-\epsilon)\mathbbm{1}/2^n + \epsilon |\psi\rangle\langle\psi|$ which captures the essential dynamics of a pure state $\ket{\psi}$. Further details of PPS preparation for two and three-qubit systems are provided in Appendix \ref{PPSseq}. 

\textit{Readout:}  The instantaneous states during evolution are read-out using full Quantum State Tomography (QST), which allows us not only to monitor populations in various energy levels, but also to quantify coherences and thereby extract quantum correlations.  Since NMR signals arise from single-transition transverse-magnetization operators of the form $I_x^i \pm i I_y^i$, not all  elements of the density matrix are directly 
measurable.  Therefore, one performs a set of experiments to systematically convert unobservable elements to observable elements of the density matrix, followed by their measurements \cite{chuang1998bulk}.  In our case, we perform six and twelve such detection experiments for two and three qubit registers respectively, to obtain pure phase absorptive signals \cite{anjusha2018optimized}, using which we reconstruct the full density matrix.

\textit{Modeling experimental imperfections:}
The two main imperfections in the NMR experiments are (i) spatial RF inhomogeneity (RFI) causing different Rabi amplitudes at different points in the sample and (ii) z-repolarization process $T_1$ (relaxation to thermal equilibrium) and dephasing process $T_2$ (loss of quantum coherence) \cite{levitt2001spin}.  The rate constants $T_1$ and $T_2$ are measured by standard NMR methods. We model RFI by considering a probability distribution of RF values spread over  $\pm 10~ \%$ about the nominal value.  This distribution is then optimized by minimizing the rms deviation of the experimental data points from the corresponding theoretical values. Decoherence effects are also incorporated into the same model. The theoretical points are obtained by solving the von Neumann equation in the rotating frame using Eq. (\ref{intH1}) for the corresponding initial state density matrix in each case. 
\begin{figure*}
	\centering
	\includegraphics[trim=0.1cm 11.5cm 0.1cm 5cm,width=18cm,clip=]{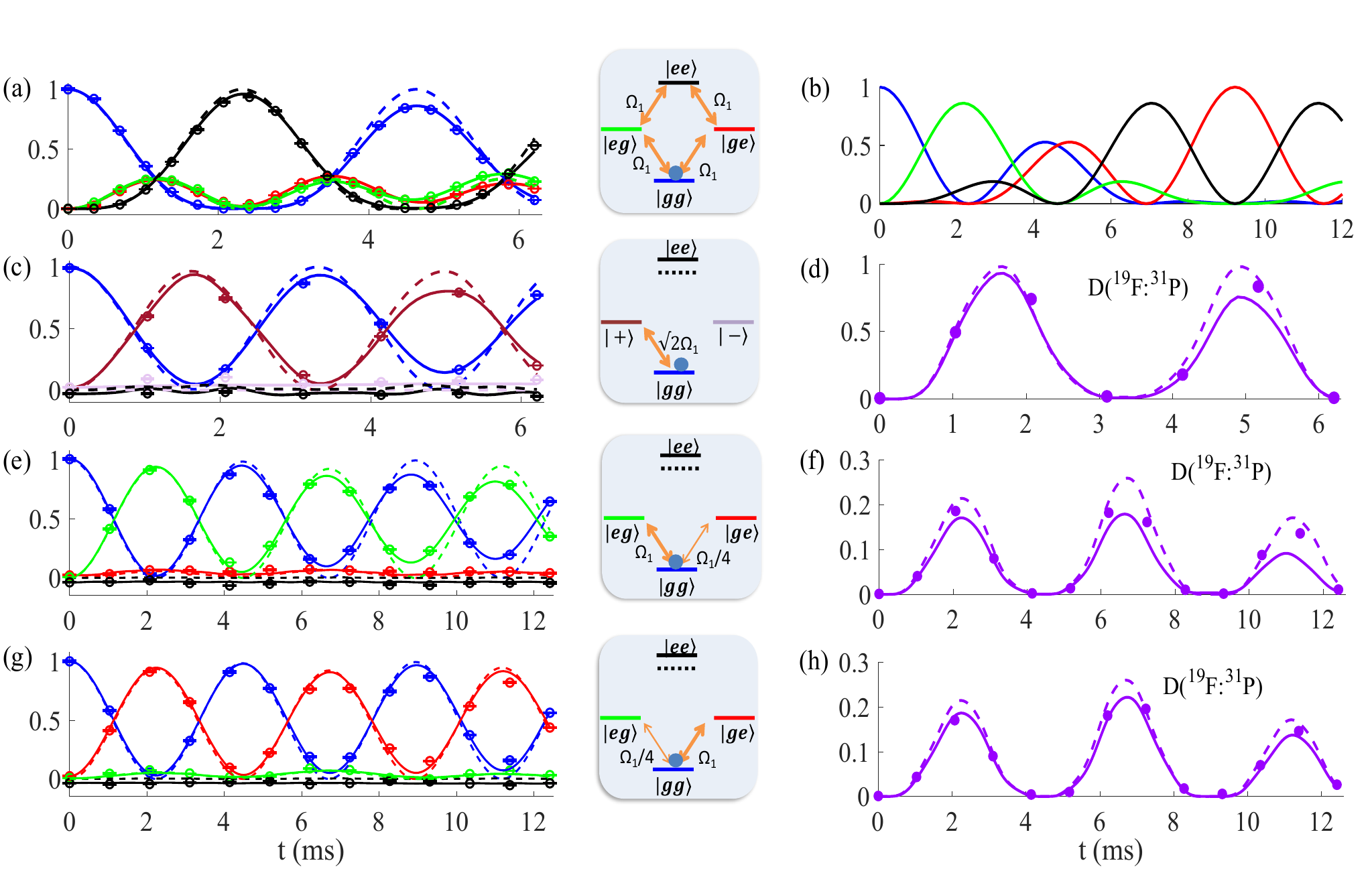}
	\caption{Population dynamics (left column) and discord (right column) versus driving time for the two-qubit register under Rabi drive (a), interaction induced blockade (c,d), freezing of the second qubit (e,f) and of the first qubit (g,h). Plot (b) shows the population dynamics of a pair of non-interacting spins with the same driving parameters as in freezing on second qubit case (g), illustrating the importance of spin-spin interaction to realize freezing. Experimental data points are shown by dots with error bars (indicating random errors), theoretically expected dynamics are shown by dashed lines, and realistic numerical models are shown by solid lines. In each case, the corresponding energy level diagram  in a relevant basis along with prominent transitions are also shown (central column) with the same color coding. The discord values $D(B|A)$ are expressed in units  of $\ln2/\epsilon^2$ \cite{katiyar2012evolution}, where $\epsilon$ is the purity factor as described in section III C.  
		}
	\label{results_2q}
\end{figure*}

\begin{figure*}
	\centering
	\includegraphics[trim=0.2cm 2.5cm 0.7cm 0.5cm,width=18cm,clip=]{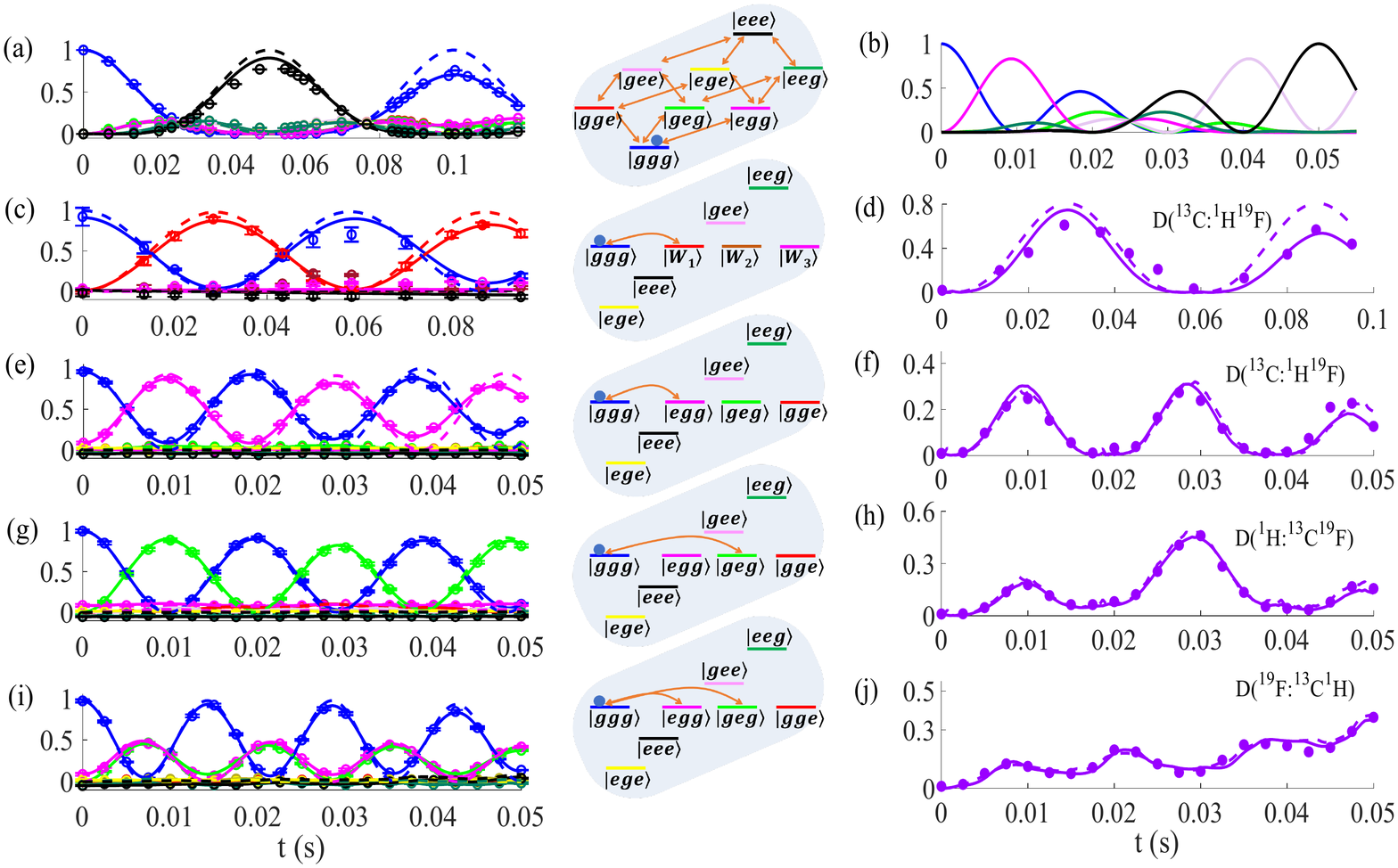}
	\caption{Population dynamics (left column) and discord (right column) versus driving time for the three-qubit register under Rabi drive (a), interaction induced blockade (c,d), freezing of the second and third qubits (e,f) and of the first and third qubits (g,h) and of only the last qubit (i,j). Plot (b) shows the population dynamics for three non-interacting spins with the same driving parameters as freezing on second and third qubits (e,f), illustrating the importance of spin-spin interactions to realize freezing. Experimental data points are shown by dots with error bars (indicating random errors), theoretically expected dynamics are shown by dashed lines, and realistic numerical models are shown by solid lines. In each case, the corresponding energy level diagram  in a relevant basis along with prominent transitions are also shown (central column) with the same color coding. The discord values $D(B|A)$ are expressed in units  of $\ln2/2\epsilon^2$ \cite{katiyar2012evolution}, where $\epsilon$ is the purity factor as described in section III C.}
	\label{results_3q}
\end{figure*}

\section{Experimental Results}
\label{NMR-Rydberg}
\subsection{Non-interacting spins ($J_{ij}=0$)}
To appreciate the interaction induced blockade and freezing effects, we first demonstrate Rabi oscillations of noninteracting ($J_{ij} = 0$) spins under a uniform drive, i.e., $ \nu_i^{\mathrm{RF}}= \nu^{\mathrm{RF}}$ for all $i$. The Zeeman energy splitting provides the necessary levels for Rabi oscillations, as illustrated in the central columns of Fig.~\ref{results_2q} and Fig.~\ref{results_3q}. The Rabi oscillation of a non-interacting spin-1/2 nucleus is studied simply using  low-bandwidth transition-selective RF fields whose carrier frequencies are set on one of the transitions of each spin and ignoring all other off-resonant transitions.  In our experiments, after initializing each spin to its ground state, as explained before, we drive the on-resonant transitions with RF fields of amplitudes $ \nu^{\mathrm{RF}} = 217 $ Hz and $\nu^{\mathrm{RF}} = 10$ Hz respectively for the two and three-qubit registers. The relative populations of ground and excited states are then measured by a suitable detection pulse after dephasing (and hence destroying) the coherences with the help of a pulsed field gradient (PFG).   

In Figs.~\ref{results_2q}(a) and \ref{results_3q}(a), we show the dynamics for the non-interacting spins in the two-qubit and three-qubit systems, respectively. We drive all the spins simultaneously, and it leads to coherent Rabi oscillations between the ground states $\ket{gg}$ or $\ket{ggg}$ with the excited states $\ket{ee}$ or $\ket{eee}$ respectively. The decay profiles indicated by the experimental data points relative to the theoretical expectations (dashed-lines) are due to environmental relaxations in NMR systems as well as RFI.  These effects are captured fairly well by the model indicated by the solid lines.
In two-qubit register, the population transfer from $|gg\rangle$ to $|ee\rangle$ takes place via single excited states $|eg\rangle$ and $|ge\rangle$ and for $N=3$ we have both singly and doubly excited states as intermediate ones as shown in Fig.~\ref{results_3q}(a).

\subsection{Strongly interacting case: Interaction induced excitation blockade}
Now, we consider the case of strongly interacting spins with $J_{ij}\gg \nu^{\mathrm{RF}}$ and $\nu_i\gg \nu^{\mathrm{RF}}$. For the two-qubit register, we have $J_{ij}= 868$ Hz, and $\nu^{\mathrm{RF}}= 217$ Hz. The corresponding dynamics is shown in Fig.~\ref{results_2q}(c) for the initial state $\ket{gg}$. We see the Rabi oscillations between $\ket{gg}$ and $\ket{+} = (\ket{ge}+\ket{eg})/\sqrt{2}$, with no population being found in $|ee\rangle$, indicating the excitation blockade. Note that, the anti-symmetric state $|-\rangle=(\ket{ge}-\ket{eg})/\sqrt{2}$ is completely decoupled from the excitation dynamics. Comparing this to the results for the non-interacting qubits [Fig.~\ref{results_2q}(a)], the oscillation frequency of the population in $|gg\rangle$ is amplified by a factor of $\sqrt{2}$ due to the blockade. Experimentally, we observe an oscillation frequency of $(\sqrt{2} \pm 0.002)\nu^{\mathrm{RF}}$, showing an excellent agreement with the expected theoretical prediction.

In the three-qubit case [see Fig.~\ref{results_3q}(c)], we have $\nu^{\mathrm{RF}}= 10$ Hz, and the interaction strengths are given in Fig.~\ref{molecules}(d). Note that one of the spin-spin couplings is negative, making $|G\rangle$ no longer the ground state in our three-qubit register in the absence of RF driving. But, $|G\rangle$ is degenerate with the Werner state $\ket{W_1}\equiv |+_3\rangle = (\ket{001}+\ket{010}+\ket{100})/\sqrt{3}$ and its orthogonal counterparts $\ket{W_2}$ and $\ket{W_3}$  in singly-excited subspace, which can be determined by Gram-Schmidt orthogonalization. The basis is not uniquely fixed, but a possible combination is $\ket{W_2} = (2\ket{001} -\ket{010}-\ket{100})/\sqrt{6}$ and $\ket{W_3} = (\ket{010}-\ket{100})/\sqrt{2}$. The spin-spin interactions are such that the states with more than one excitation are energetically well separated from $|G\rangle$ and singly excited states. The experimental and theoretical results of the population dynamics shown in Fig. \ref{results_3q}(c) for the initial state $|G\rangle$ indicate that the population exchange occurs only between $\ket{G}$ and $\ket{W_1}$, while all other states are blocked.  Here again the collective Rabi oscillation has a frequency of $(\sqrt{3} \pm 0.03)\nu^{RF}$, which also shows excellent agreement with the expected value $\sqrt{3}\nu^{RF}$.

Figs.~\ref{results_2q}(d) and \ref{results_3q}(d) show the quantum discord between A:B and X:YZ at different steps of evolution under blockade conditions for two and three-qubit cases respectively. Initially, the system is prepared in a product state $|G\rangle$, and hence the quantum discord is zero. However, during the course of time evolution, correlations develop between the qubits, resulting in non-zero values of discord. We can see that the discord is maximized each time when the system attains the entangled state $|+\rangle$ in the case of two-qubits and $\ket{W_1}$ in the case of three-qubits. After incorporating the imperfections, the numerical model (solid lines) and experimental (circles) results are in excellent agreement.


\subsection{Strongly interacting case: Local spin freezing}
In the blockade regime, by locally amplifying the Rabi coupling (or equivalently the local transverse field in the spin model in Eq.~(\ref{intH1})) of selected spins, we can freeze the dynamics of other spins, which in a Rydberg lattice is called the Rydberg biased freezing \cite{sri19}. In the two qubit NMR register, we drive the first qubit ($^{19}F$) with $\nu_1^{\mathrm{RF}} = 217$ Hz and the second qubit ($^{31}P$) with a weaker field, i.e., $\nu_2^{\mathrm{RF}} = \nu_1^{\mathrm{RF}}/4$. The corresponding dynamics is shown in Fig.~\ref{results_2q}(e). For these values of field and interaction strengths, we expect freezing of the second qubit and it remains in its ground state $|g\rangle$, and the two-qubit system exhibits Rabi oscillations between $|G\rangle$ and $|eg\rangle$ as shown by dashed lines in Fig.~\ref{results_2q}(e). Due to the imperfections discussed in Sec.~\ref{imp}, we experimentally observe a small fraction of population in $|ge\rangle$ (circles in Fig.~\ref{results_2q}(e)). After incorporating the imperfections, numerical model results (solid lines) show excellent agreement with the experimental values. To appreciate the biased spin freezing due to the strong spin-spin interactions and the inhomogeneous Rabi coupling, we show the same dynamics as that of the non-interacting qubits, but with $\nu_2^{\mathrm{RF}} = \nu_1^{\mathrm{RF}}/4$. In the latter case, we can see that both qubits are involved in the excitation dynamics [see Fig.~\ref{results_2q}(b)]. If we switch the weaker drive to the first qubit and the stronger one to the second qubit, we observe prominent dynamics of the second qubit while the first qubit freezes in the presence of strong spin-spin interactions, as shown in Fig.~\ref{results_2q}(g). This effect is persistent for 12 ms, which is about 0.65 times the time period of the weaker drive. These results also show that this behaviour is not transient, but holds over extended times of evolution.  

\begin{figure}[h!]
	\centering
	\includegraphics[trim=0.8cm 1cm 0.5cm 1.5cm,width=9cm,clip=]{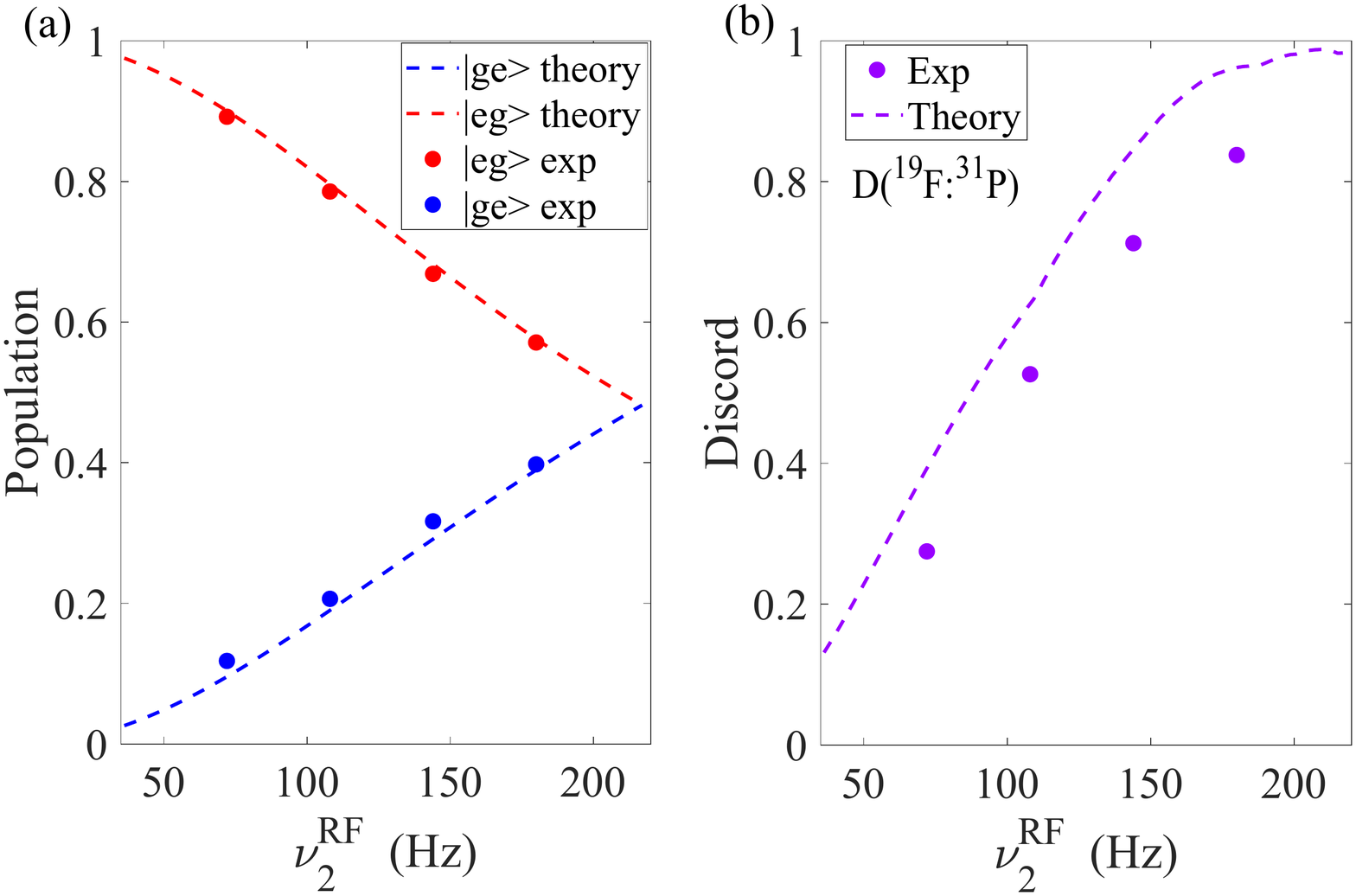}
	\caption{(a) Population in $\ket{ge}$ and $\ket{eg}$ as the driving amplitude of the second qubit ($\nu_2^{RF}$) is gradually reduced from Rydberg blockade condition ($\nu_1^{RF} = 217$ Hz) to Rydberg biased freezing condition (54.2 Hz) (b) the corresponding discord values $D(A|B)$ in units of $\ln2/\epsilon^2$. Experimental data are recorded at half the effective Rabi period (for each value of $\nu_2^{RF}$) is shown by filled circles, which are overlaid on theoretical simulations shown by dashed lines.}
	\label{blk2frz}
\end{figure}

In addition to this, we also studied the regime in between Rydberg blockade and Rydberg biased freezing by gradually reducing the driving amplitude of the second qubit with respect to the first qubit and observed the populations in each of the singly excited states $\ket{ge}$ and $\ket{eg}$. In Fig.~\ref{blk2frz}(a), we show the value of population in states $|eg\rangle$ and $|ge\rangle$ at half the effective Rabi period, i.e., at the first crest that appeared in the Rabi dynamics. We can see that the populations start out equally in $\ket{ge}$ and $\ket{eg}$ states in blockade regime, with driving amplitude 217 Hz on both qubits and gradually deviate from each other as the driving fields on both qubits become different. Here, the driving amplitude of the second qubit is reduced, and hence the population in $\ket{ge}$ gradually decreases while the population in $\ket{eg}$ increases.  The corresponding quantum discord values for each of these points is shown in Fig.~\ref{blk2frz}(b). We can see that as the driving amplitude of the second qubit is increased, the discord also increases, signalling the shift from freezing to blockade regime. This is expected, since with increasing drive amplitude, the system dynamics is no longer restricted to the subspace of the first qubit. The second qubit dynamics also become prominent, finally resulting in the maximally entangled $\ket{+}$ state under blockade condition.

In the three-qubit register, we can selectively freeze either a single qubit or two qubits. To demonstrate two-qubit freezing we drive the first qubit with $\nu_1^{\mathrm{RF}} = 50$ Hz and the last two quibits by $\nu_3^{\mathrm{RF}} = \nu_2^{\mathrm{RF}}=10$ Hz. As seen in Fig.~\ref{results_3q}(e), the first qubit only takes part in the excitation dynamics, resulting in Rabi oscillations between $|G\rangle$ and $|egg\rangle$. Instead of the first qubit, if we drive the second qubit strongly, and weakly drive the first and the third ones, we observe Rabi oscillations between $|G\rangle$ and $|geg\rangle$ [see Fig.~\ref{results_3q}(g)]. To demonstrate single qubit freezing in the three qubit register, we drive the third qubit weakly in comparison to the first two qubits, i.e., $\nu_1^{\mathrm{RF}} = \nu_2^{\mathrm{RF}} = 50$ Hz, $\nu_3^{\mathrm{RF}} = 10$ Hz. As shown in Fig.~\ref{results_3q}(i), the population dynamics in this case is between $|G\rangle$ and the single excitation states of the first two qubits, $|egg\rangle$ and $|geg\rangle$ respectively, while the third qubit dynamics is suppressed. Similar to the two-qubit case, in the absence of spin-spin couplings between the qubits and under non-uniform drive $\nu_2^{\mathrm{RF}} = \nu_3^{\mathrm{RF}} = \nu_1^{\mathrm{RF}}/5$, all qubits get excited simultaneously, as shown in Fig. \ref{results_3q}(b). This reinforces the fact that strong spin-spin interactions cause local spin freezing and that this phenomenon sustains over extended periods of evolution under such conditions.

Figs.~\ref{results_2q}(f,h) and \ref{results_3q}(f,h,j) show discord between A:B and X:YZ for Rydberg biased freezing scenarios in two and three qubit cases respectively. In the three-qubit case, discord is calculated for different partitions as indicated in the figure \ref{results_3q}(f,h,j). We can see that less entanglement is generated under conditions of Rydberg biased freezing as compared to Rydberg blockade. This is due to considerable suppression of the dynamics of frozen qubits during the evolution. Accordingly, the dynamics is largely confined to exchanges between separable states, with less quantum correlation being created as revealed by discord values. We can also see in Fig.~\ref{results_3q}(j) that discord of the frozen qubit with the rest of the system is steadily increasing. 
Further weakening of the drive amplitude on the frozen qubit will lead to its stronger isolation and further suppression of quantum correlations.

%
%

\section{Summary and Outlook}
\label{Conclusions}
In this work, we experimentally demonstrated interaction induced blockade and local spin freezing using two and three-qubit NMR registers. Both the phenomena are identified and studied in the context of Rydberg atoms, and are known as Rydberg blockade and Rydberg biased freezing. While Rydberg blockade has previously been demonstrated experimentally, we believe this is the first experimental demonstration of local spin freezing which also simulates Rydberg biased freezing.
 In addition, we have also characterized the dynamics of quantum discord in these systems during the course of evolution under blockade and spin freezing conditions. 

Though, the concepts of NMR spin systems are altogether different from that of Rydberg atoms, it does not hinder access to quantum phenomena featured by the latter, thus justifying the role of NMR registers as versatile quantum simulators.

The ability of interaction induced blockade, realized by a simple off-resonant drive, to create multi-qubit entanglement may have interesting applications in experimental quantum information studies.  The robustness of such approaches compared to the traditional methods involving a combination of local and nonlocal gates is hitherto unexplored.  
Moreover, the biased local spin freezing can be utilized to selectively control and drive qubits in a multi-qubit system. This may find applications for local quantum control,  such as exchanging information among a subset of qubits and controlled generation of quantum dynamics in a system of interacting qubits.

Finally, the operational maps between atoms and spins used in these studies might open up new possibilities in experimental quantum simulations using NMR architecture.  Such maps not only provide clues towards new phenomena not foreseen by either of the quantum architectures, but also help conceive hybrid architectures.

\section*{Acknowledgments}
We acknowledge useful discussions with Ankita Niranjan and Vineesha Srivastava. R. N. acknowledges funding from UKIERI- UGC Thematic Partnership under Grant No. IND/CONT/G/16-17/73.  TSM acknowledges funding from DST/ICPS/QuST/2019/Q67.

\appendix
\section{Preparation of two and three-qubit PPS}

\label{PPSseq}
\begin{figure}[h!]
	\includegraphics[trim=0cm 9.5cm 0cm 0cm,clip=,width=8cm]{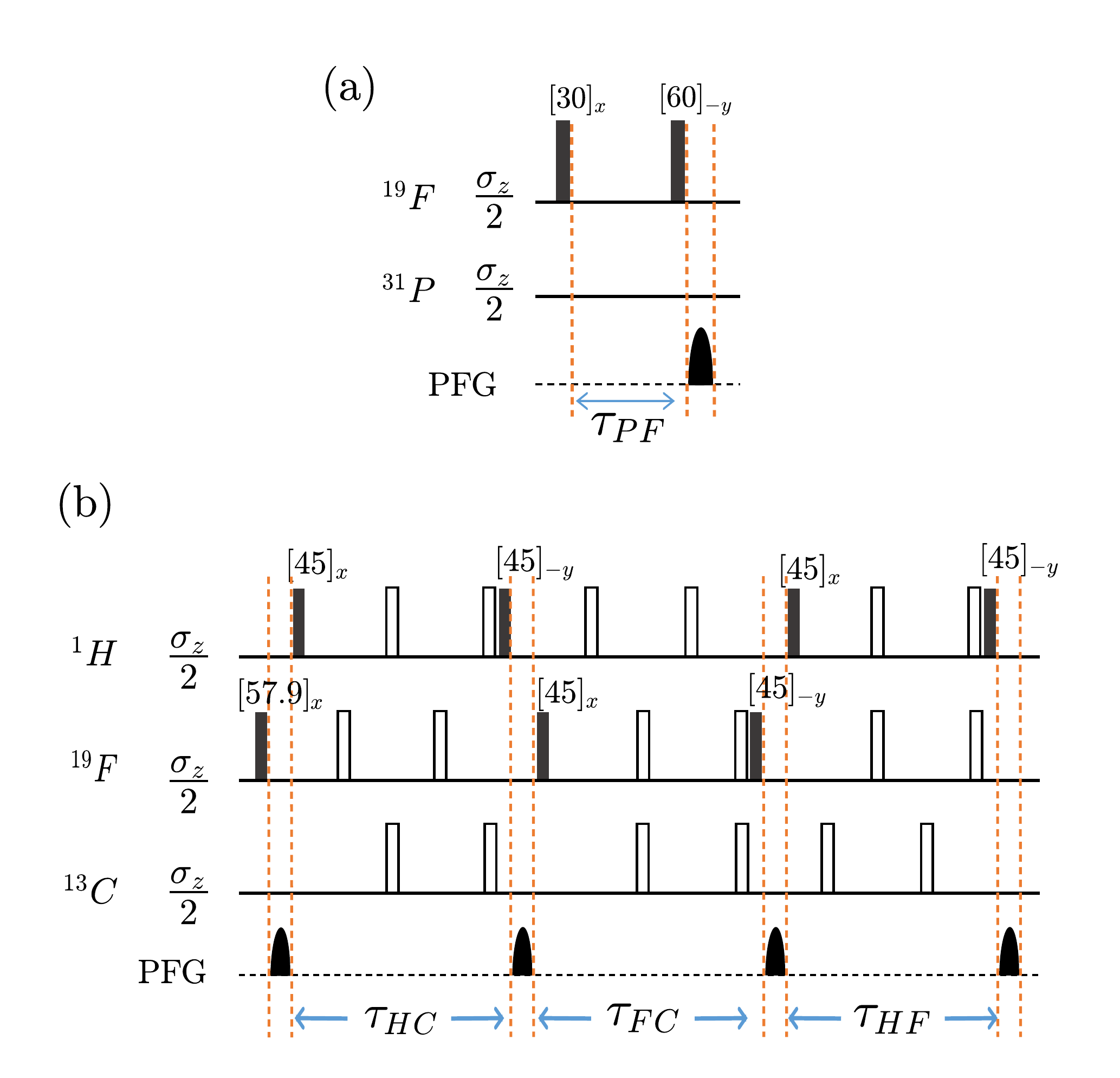}
	\caption{PPS sequences. Pulse sequence for preparing PPS $|00\rangle\langle00|$ and $|000\rangle\langle000|$ from thermal equilibrium state for (a) two qubit register in sodium fluorophosphate molecule with $\tau_{FP} = 1/(2J_{FP})$ and (b) three qubit register in dibromofluoromethane with $\tau_{HC} = 1/(2J_{HC})$, $\tau_{FC} = 4(1/J_{FC} - 1/(8J_{HC}))$ and $\tau_{HF} = 1/(2J_{HF})$  respectively. The solid bars represent rotations by an angle and about a direction as indicated over them. The blank rectangles represent $\pi$ pulses and the black half ellipsoids represent PFG along $+$z axis to destroy coherences.}
	\label{PPS}
\end{figure}

As explained in the main text, at ambient temperatures, in NMR spin systems, the Zeeman energy gaps are much smaller than the thermal energy. Hence, there is very little population difference between the energy levels, resulting in highly mixed states. To obtain pure states, very low temperatures or unrealistically high magnetic fields are required. This shortcoming can be overcome by preparing the systems in a pseudopure state (PPS), in which state, the populations are equal in all energy levels except one level of interest. The dynamics of a PPS can be mapped isomorphically to that of pure states \cite{cory1997ensemble}. The pulse sequence to prepare $\ket{00}$ PPS ($\ket{gg}$) in two-qubit system sodium fluorophosphate molecule is shown in Fig. \ref{PPS} (a), and $\ket{000}$ PPS ($\ket{ggg}$) in three-qubit system dibromofluoromethane system is shown in Fig. \ref{PPS} (b). 

\bibliographystyle{apsrev4-1}
\bibliography{references}
  \end{document}